\documentstyle[preprint,revtex]{aps}
\math-with-secnums
\begin{document}
\draft
\begin{title}
The $^3S_1-^3D_1$ effective-range parameters of some realistic nucleon-nucleon
potentials
\end{title}
\author{Wytse van Dijk}
\begin{instit}
Redeemer College, Ancaster, Ontario, Canada L9G 3N6  \\
and Department of Physics and Astronomy, McMaster University,  \\
Hamilton, Ontario, Canada L8S 4M1
\end{instit}
\moreauthors{Mark W. Kermode}
\begin{instit}
Department of Applied Mathematics and Theoretical Physics,  \\
University of Liverpool, PO Box 147, Liverpool, UK  L69 3BX
\end{instit}
\moreauthors{Dao-Chen Zheng\cite{address}}
\begin{instit}
Department of Physics and Astronomy, McMaster University, \\
Hamilton, Ontario, Canada L8S 4M1
\end{instit}
\receipt{November 20, 1992, accepted Phys. Rev. C}

\newpage
\begin{abstract}
We consider the $P$- and $Q$-shape parameters in the eigenphase and
`bar' phase shift representations for the two Reid potentials and the
more recent Moscow potential.  The values of these parameters were
obtained from integrals of the zero-energy wave function in a similar
manner to that for the effective range.  Such expressions for the
shape parameters yield more reliable numerical values of the shape
parameter than those obtained by least-mean-square or polynomial fits.
We show that the eigenphase representation is more meaningful than the
`bar' phase shift representation and that the variation in the values
of $P_\alpha$ in the models may be of significance.
\end{abstract}


\newpage

\narrowtext

\section{Introduction}

The low-energy scattering phase shifts are conveniently described by an
effective-range expansion,
\begin{equation}
y(k^2) \equiv k\cot(\delta) = -1/a + (1/2)r_0k^2 - Pr_0^3k^4 + Qr_0^5k^6 +
\cdots.   \label{1}
\end{equation}
Although the scattering length $a$ and the effective range $r_0$ are
quantities that can be determined quite precisely from experiment, the
coefficients of the higher order terms in $k^2$ are less amenable to
experimental evaluation.  Nevertheless, the shape parameter $P$ (and
possibly also $Q$) is of interest in comparing realistic models of
the nucleon-nucleon interaction.  Furthermore, it has recently been
shown that the shape parameter plays a role in the expansion of the
ratio of the deuteron radius $r_D$ to the triplet scattering
length $a_t$ in terms of a variable $x$, which is proportional to the square
root of the binding
energy $\alpha $ \cite{kermode91}, {\it i.e.,}
\begin{equation}
\sqrt 8 r_D/a_t = 1 + (1/2) x^2 + a_3 x^3 + a_4 x^4 + \cdots,
\end{equation}
where $x=(1/2) \alpha r_0$.
(Related expansions derived earlier \cite{bhaduri90,sprung90} do not show
the explicit dependence on the shape parameter.)
The higher order coefficients ($a_3, \cdots$) of the expansion
depend on the shape parameter, for example,
$a_3 = 1 - 2J + 8P$, where $J$ involves integrals containing the
zero-energy wave functions.
If we substitute the appropriate value for the Reid soft core
potential \cite{reid68} in this expression, the left hand side
takes the value 1.0269 and the {\it first two} terms on the right
hand side give 1.0199.  Thus the determination of $a_3$, and perhaps
$a_4$, along with suitable modification of the expression for coupled
channels, should provide a rather accurate formula for the deuteron
radius in terms of the scattering parameters.
In this paper we concentrate on one component of $a_3$ or $a_4$,
namely the shape parameters $P$ and $Q$ respectively.

The accurate determination of $P$ (and $Q$) from realistic
nonrelativistic nuclear potentials is not as trivial a task as it
might appear.  For the same potential model, significantly varying
values for the shape parameter are quoted by different
authors\cite{reid68,kermode81,kermode91a,wong92,vandijk92}.  In Table I
we give a sample of values of the shape parameters for the Reid
hard-core potential.  The shape parameters are usually determined by
performing a least-mean-square fit of $k \cot(\delta)$ to a
polynomial of specified degree in $k^2$, or a simple polynomial fit to
a polynomial of degree $n-1$ in $k^2$ where $n$ is the number of (low)
energies at which the phase shifts have been
evaluated\cite{reid68,kermode81,wong92}.

The triplet nucleon-nucleon scattering matrix can be described in terms
of different kinds of scattering phase shifts, the most common of
which are the eigenphases\cite{blatt52} and the `bar' phase
shifts\cite{stapp57}.  Whereas the scattering length and the effective
range are the same for the eigenphases and `bar' phase shifts, the
corresponding shape parameters are different for the two cases.  This
difference is not always made clear, leading to ambiguities.
Wong\cite{wong92}, for example, seems to be concerned with eigenphases
in his analysis, but compares his value of the shape parameter
for the Reid hard-core potential to that given by Reid\cite{reid68},
who calculated the shape parameter from the `bar' phase shifts.  In a
recent paper\cite{vandijk92} we derived a relationship between the two
shape parameters, which shows that the difference is a simple
expression of measurable quantities, {\it i.e.,}
\begin{equation}
P_b - P_e = \frac{q^2}{a_t r_0^3},   \label{2}
\end{equation}
where $a_t$ and $r_0$ are the triplet scattering length and effective
range respectively. The quantity $q$ is the zero-energy slope of the
function $f(k^2) = \tan \epsilon$, $\epsilon$ being the eigen-mixing
parameter.  We also showed that the shape parameter $P_e$ can be
expressed in terms of an integral involving the zero-energy wave
function.

 In this paper we show that one can make use of the integral
formulation to obtain more reliable numerical values of the shape
parameter $P$ and the next coefficient in the effective-range
expansion $Q$.  In Sec. II we show the relationship between the
eigenphase and the `bar' phase shift effective-range parameters.  The
parameters may be determined from the zero-energy wave function by the
method described in Sec. III.  In Sec. IV we report some numerical
results.

\section {Effective-range expansions}

In this section we obtain a formula for the difference between the
eigenphase and `bar' phase shift shape parameters $P$ and $Q$ by using
the relationships between the eigenphases and `bar' phase shifts.

Extending the effective-range expansions for the eigenphases
given by Biedenharn and Blatt\cite{biedenharn54}, and using their
notation, we obtain
\begin{equation}
k \cot \delta_\alpha = -\frac{1}{a_\alpha} + \frac{1}{2} r_\alpha k^2
- P_\alpha r_\alpha^3 k^4 + Q_\alpha r_\alpha^5 k^6 - \cdots,  \label{3}
\end{equation}
\begin{equation}
k^5 \cot \delta_\beta = \frac{225}{a_\beta^4}\left[ -\frac{1}{a_\beta}
+ \frac{1}{2}r_\beta k^2 - P_\beta r_\beta^3 k^4 + Q_\beta r_\beta^5
k^6 - \cdots \right],  \label{4}
\end{equation}
and
\begin{equation}
\tan \epsilon = q k^2 + q_1 k^4 + q_2 k^6 + \cdots.  \label{5}
\end{equation}
The quantities subscripted with $\alpha$  or $\beta$ correspond to the state in
which the $S$ or $D$ wave respectively dominates.
The eigenphases and `bar' phase shifts are related by the
equations\cite{mott65}
\begin{equation}
\bar \delta_1  + \bar \delta_2 = \delta _\alpha + \delta _\beta,
\label{6}
\end{equation}
\begin{equation}
\bar \delta_1  - \bar \delta_2 = \arcsin (\tan 2 \bar {\epsilon}
/ \tan 2 \epsilon),  \label{7}
\end{equation}
and
\begin{equation}
\bar \epsilon = {1\over 2} \arcsin [\sin 2 \epsilon
\sin (\delta _\alpha - \delta _\beta)] . \label{8}
\end{equation}
Using these equations, we derive an effective-range expansion for
$\bar \delta_1$ in
terms of the eigenphase parameters.  Thus we obtain
\begin{equation}
k \cot \bar \delta_1 = -\frac{1}{a_\alpha} + \frac{1}{2} r_\alpha k^2
- \left(P_\alpha r_\alpha^3 + \frac{q^2}{a_\alpha}\right) k^4
+  \left( Q_\alpha r_\alpha^5 + \frac{1}{2} q^2 r_\alpha -
\frac{2qq_1}{a_\alpha} \right) k^6 + \cdots.  \label{9}
\end{equation}
In this expansion we identify the `bar' phase shift shape parameters $\bar
P_1$ and $\bar Q_1$.  The scattering length is the same for both
parameterisations and so is the effective range.  The differences
between the `bar' phase shift
and the corresponding eigenphase shape parameters are simple expressions,
\begin{equation}
\bar P_1 -P_\alpha = \frac{q^2}{a_\alpha r_\alpha^3}  \label{10}
\end{equation}
and
\begin{equation}
\bar Q_1 - Q_\alpha = \frac{q(qr_\alpha a_\alpha - 4 q_1)}{2 a_\alpha
r_\alpha^5}.  \label{11}
\end{equation}
Note that we have now written $P_\alpha$ and $\bar P_1$ for $P_e$
and $P_b$ of Eq. (1.2).

\section{Effective-range parameters from zero-energy
wave function}

   From the previous section it is evident that given the eigenphase
effective-range parameters we can determine the `bar' phase shift
effective-range parameters.  In this section we show that the
eigenphase effective-range parameters, including the shape parameters
and $q$ and $q_1$, can be found from the zero-energy wave function,
and that this way of determining them leads to numerically more
precise, and less ambiguous, results than the conventional method of
fitting polynomials to low-energy values of $k \cot \delta$.

The zero-energy wave function satisfies the Schr\"odinger equation,
\begin{equation}
-u''_0 + V_C u_0 + 2\sqrt 2 V_T w_0 = 0,   \label{12}
\end{equation}
\begin{eqnarray}
-w''_0 & + & V_C w_0 + 6r^{-2} w_0 \nonumber  \\
       & - & 2 V_T w_0 - 3 V_{LS} w_0
+ 2\sqrt 2 V_T u_0 = 0.   \label{13}
\end{eqnarray}
We determine the zero-energy eigenphase solutions, whose asymptotic form
({\it i.e.,} for $r > R$, the  nuclear force range) is
\begin{eqnarray}
\psi_{0\alpha}(r)
= \left( \begin{array}{c}
                        u_{0\alpha}  \\  w_{0\alpha}
                        \end{array}        \right)
\sim  \left( \begin{array}{c} \tilde u_{0\alpha} \\ \tilde w_{0\alpha}
\end{array}   \right)
 & =  & \left( \begin{array}{c}
            1 - r/a_\alpha  \\  3q/r^2
            \end{array}       \right),        \label{14}   \\
\psi_{0\beta}(r) = \left( \begin{array}{c}
                        u_{0\beta}  \\  w_{0\beta}
                        \end{array}        \right)
\sim   \left( \begin{array}{c}
                      \tilde  u_{0\beta}  \\  \tilde w_{0\beta}
                        \end{array}        \right)
& = & \left( \begin{array}{c}
            15qr  \\  -r^3 + a_\beta^5/5 r^2
            \end{array}       \right),        \label{14.1}
\end{eqnarray}
Except for the normalization, these solutions are given in reference
\cite{biedenharn54}.  To obtain these solutions numerically, one can
integrate four different solutions ($\phi_i,i=1,\dots,4$) outward and
match them at $R$ to
obtain the appropriate asymptotic form.  Since asymptotically there
are terms which become larger ($r^3$ term) and smaller ($r^{-2}$
term) as $r$ increases, we have found that inward integration gives
more precise
results.  In this case one starts integrating inward to the potential
hard core radius $c$ from the asymptotic
region, say $r=R$, where
\begin{equation}
\phi_1 = \left( \begin{array}{c}  1 \\ 0  \end{array} \right),
\phi_2 = \left( \begin{array}{c}  r \\ 0  \end{array} \right),
\phi_3 = \left( \begin{array}{c}  0 \\ r^{-2}  \end{array} \right),
\phi_4 = \left( \begin{array}{c}  0 \\ -r^3  \end{array} \right).
\label{15}
\end{equation}
The eigenphase solutions are then
\begin{equation}
\psi_{0\alpha} = \phi_1 + t_2^{(\alpha)} \phi_2 + t_3^{(\alpha)} \phi_3
\label{15.1}
\end{equation}
and
\begin{equation}
\psi_{0\beta} = \phi_4 + t_2^{(\beta)} \phi_2 + t_3^{(\beta)}
\phi_3.   \label{16}
\end{equation}
The condition that
\begin{equation}
\psi_{0\sigma}(c) = \left( \begin{array}{c} 0 \\ 0 \end{array}
\right), \hspace{.2in}  \mbox{ where } \sigma = \alpha \mbox{ or } \beta,
 \label{17}
\end{equation}
and where $c$ is the potential hard-core radius or zero if the potential
has no hard core,
permits us to determine the constants $t_{2,3}^{(\alpha,\beta)}$.
These constants will yield the scattering lengths
\begin{equation}
a_\alpha = -1/t_2^{(\alpha)} \hspace{.2in} \mbox{and} \hspace{.2in}
a_\beta^5 = 5 t_3^{(\beta)},  \label{18}
\end{equation}
and the quantity
\begin{equation}
q=t_3^{(\alpha)}/3 = t_2^{(\beta)}/15.   \label{19}
\end{equation}
The last equation may be used as a check of the numerical consistency
of the calculation.

The quantities $q_1$, $r_\alpha$, and $P_\alpha$ are expressed in terms
of integrals of the zero-energy wave functions and their asymptotic
forms given in Eqs. (\ref{14}) and (\ref{14.1}).  The expression for
$q_1$ is derived in reference \cite{biedenharn54} and, with an
adjustment for a change in normalization, is
\begin{equation}
q_1 = \frac{1}{15} \int_0^\infty [\tilde u_{0\alpha} \tilde u_{0\beta} -
3qr - u_{0\alpha} u_{0\beta} - w_{0\alpha} w_{0\beta} ] dr.  \label{20}
\end{equation}
We can express this integral as one in which one needs to integrate numerically
 only over the range of distances over which the potential acts; {\it
i.e.,} for $R$ as the range of interaction,
\begin{equation}
q_1 = \frac{1}{15} \left\{ q R^2 \left(6-5\frac{R}{a_\alpha}\right) -
\int_c^R (u_{0\alpha} u_{0\beta} + w_{0\alpha} w_{0\beta} ) dr -
\frac{a_\beta^5 q}{5R^3} \right\},  \label{21}
\end{equation}
where $c$ is the hard core radius of the potential; it is set equal to
zero if the potential has no hard core.

The effective range is given by the equation
\begin{eqnarray}
r_\alpha & = & 2 \int_0^\infty (\tilde u_{0\alpha}^2 - u_{0\alpha}^2 -
 w_{0\alpha}^2)
dr    \label{21.1}  \\
  & = & 2R \left( 1 - \frac{R}{a_\alpha} + \frac{R^2}{3 a_\alpha^2}
\right) - 2 \int_c^R (u_{0\alpha}^2 + w_{0\alpha}^2) dr - 6
\frac{q^2}{R^3}.     \label{22}
\end{eqnarray}
In Eqs. (\ref{21}) and (\ref{22}) the terms with positive powers of
$R$ are cancelled by contributions of the integral when $R$ approaches
infinity.
In order to evaluate the shape parameter $P_\alpha$ in terms of an
integral of the zero-energy wave function, we need the derivative of
the zero-energy wave function with respect to the
energy\cite{vandijk92,kermode90}.  We obtain the latter by expanding
the wave function $\psi_{\alpha}$ in powers of $k^2$, {\it i.e.,}
\begin{equation}
u_\alpha(r) = u_{0\alpha}(r) + k^2 u_{2\alpha}(r) + \cdots,
\label{23}
\end{equation}
\begin{equation}
w_\alpha(r) = w_{0\alpha}(r) + k^2 w_{2\alpha}(r) + \cdots.
\label{24}
\end{equation}
The functions $u_{2\alpha}$ and $w_{2\alpha}$ are then solutions of
the system of inhomogeneous equations\cite{biedenharn54,vandijk92},
\begin{equation}
-u''_{2\alpha} + V_C u_{2\alpha} + 2\sqrt 2 V_T w_{2\alpha} =
u_{0\alpha},  \label{25}
\end{equation}
\begin{eqnarray}
-w''_{2\alpha}  & + & V_C w_{2\alpha} + 6r^{-2} w_{2\alpha}   \nonumber  \\
        & - & 2 V_T w_{2\alpha} - 3 V_{LS} w_{2\alpha} +
   2\sqrt 2 V_T u_{2\alpha} = w_{0\alpha},       \label{26}
\end{eqnarray}
with the boundary conditions that for $r > R$,
\begin{eqnarray}
u_{2\alpha} \sim \tilde u_{2\alpha} & = & \frac{1}{2} r_\alpha r -
\frac{1}{2} r^2 + \frac{r^3}{6a_\alpha},   \label{27}  \\
w_{2\alpha} \sim \tilde w_{2\alpha} & = & \frac{1}{2} q +
\frac{3q_1}{r^2}. \label{28}
\end{eqnarray}
When the functions $u_{2\alpha}$ and $w_{2\alpha}$ are found
numerically, one obtains greater accuracy by integrating inward from
the asymptotic region.  The shape parameter $P_\alpha$ is given by
\cite{vandijk92}
\begin{eqnarray}
P_\alpha r_\alpha^3 & = & \int_c^R (u_{0\alpha} u_{2\alpha} + w_{0\alpha}
w_{2\alpha}) dr  - R^2 \left[ \frac{1}{4} r_\alpha - \frac{1}{6}
\left( 1 + \frac{r_\alpha}{a_\alpha} \right) R + \frac{R^2}{6a_\alpha}
- \frac{R^3}{30a_\alpha^2} \right]  \nonumber   \\
& & - \frac{q^2}{a_\alpha} + 3q
\left( \frac{q}{2R} + \frac{q_1}{R^3} \right).   \label{29}
\end{eqnarray}

One could extend the derivation of the $P_\alpha$ shape parameter in
terms of the zero-energy wave function  to
obtain the next shape parameter $Q_\alpha$ in terms of integrals of
the zero-energy wave function.  In this case one would have to obtain
the second derivative of the zero-energy wave function with respect to
energy as the solution of another inhomogeneous system of
differential equations.  Rather than introduce this complication, we
have found that a reasonably accurate value of $Q_\alpha$ can be found
by the following methods.  Supposing that $a_\alpha$, $r_\alpha$, and
$P_\alpha$ have been found, we define the function
\begin{equation}
h(k^2) \equiv \left(k \cot \delta_\alpha + \frac{1}{a_\alpha} - \frac{1}{2}
r_\alpha k^2 + P_\alpha r_\alpha^3 k^4\right)/k^6.   \label{29.1}
\end{equation}
This function is a power series in $k^2$.  We can fit a polynomial of
$k^2$ or a rational function of $k^2$ and extrapolate
 to the value for
$k^2=0$ to yield $Q_\alpha r_\alpha^5$ using the numerical recipes
POLINT and RATINT of Press {\it et al.} \cite{press89}.  Alternatively, we
 define
\begin{equation}
g(k^2) \equiv \left(k \cot \delta_\alpha + \frac{1}{a_\alpha} - \frac{1}{2}
r_\alpha k^2 + P_\alpha r_\alpha^3 k^4\right)/k^4.   \label{30}
\end{equation}
This function is a power series in $k^2$, the first term of which is
$Q_\alpha r_\alpha^5 k^2$.  Thus $Q_\alpha$ is obtained from the
relation
\begin{equation}
\left. \frac{dg}{dk^2}\right|_{k^2=0} = Q_\alpha r_\alpha^5.  \label{31}
\end{equation}
Numerically this derivative can be obtained using the five- or
six-point formulas for the derivative when $g(k^2)$ is evaluated at
equal intervals in $k^2$\cite{abramowitz65}.

\section{Results}

The shape parameters of some realistic nucleon-nucleon potentials were
calculated. These were the Reid hard and soft core potentials
\cite{reid68} and the Moscow potential \cite{kukulin89}.
The latter potential fits the experimental scattering parameters and
the deuteron binding energy, but has an extra bound state at 442 MeV.
It was included to see whether the extra node in the S-state
wave function caused any numerical problems with the integrals for
$r_\alpha$ (\ref{22}) and $P_\alpha$ (\ref{29}).  [Nonlocal potentials
without an extra bound state may have similar wave functions
\cite{kermode91b}].  We found that although the numerical calculations
are straightforward, a little care was required in the calculations of
the $Q$-shape parameter.  The second method (defining the function
$g(k^2)$ as in Eq. (\ref{30})) gave an unambiguous result, whereas the
fourth decimal place depends on the order of the polynomial when
either the polynomial (POLINT) or rational function (RATINT) fit was
used with the function $h(k^2)$ given in Eq. (\ref{29.1}).  Both
methods give results of superior precision compared to
least-mean-square or polynomial fits of the effective range expansion.
The optimum value for the energy interval used for the polynomial or
derivative fit of $Q$ is approximately 1.1 MeV.

  The results are summarized in Table II.  With the exception of the
$P$-shape parameter, the effective-range parameters of the various
realistic interactions are fairly similar.  The $P$-shape parameter has
substantially different values for different potential, but at present
we cannot suggest a physical reason for this.  The magnitude of
$P_\alpha$ is by far the largest for the Reid soft-core potential.
Whether this is a characteristic feature of the soft-core of the
potential - the `superdeep' Moscow potential, in some respects behaves
like a hard-core potential - will be investigated in the future by
considering a larger and more varied sample of realistic potentials.

   It is interesting to note that the magnitude of the $P$-shape
parameter is much smaller than the $Q$-shape parameter.  This
indicates that the radius of convergence of the effective-range series
is small. It is well known to be restricted by the one-pion-exchange
cut ($\infty < k^2 \le - \mu^2/4 \approx 0.123$) (see, for example
\cite{kermode84a}). However, for determining the deuteron radius from
the scattering data this is not too important for the eigenphase case,
since $\vert \alpha^2 Q_\alpha /P_\alpha) \vert < 0.5$ so the term in
$Q_\alpha$ may be neglected.  However, for the `bar' phase shift case,
$\vert \alpha^2 \bar Q_1 /\bar P_1 \vert $ can be significant (1.9 in
the case of the Moscow potential) so $\bar Q_1$ is not unimportant.
This is related to the fact that the `bar' effective-range function
has infinite derivative at the deuteron bound state
\cite{kermode83a}.  The results for the Reid hard-core potentials with
different potential ranges show that the shape parameters are not
greatly affected by the long-range tail of the potential.

Although experimental values for the shape parameter do not exist, it
would be interesting to know their values from the smooth curves
through the experimental data obtained from phase shift analyses.
This may be helpful in the comparison of particular models.

\nonum
\section{Acknowledgments}

One of us (M.W.K.)  thanks the members of the Theoretical Physics
group at McMaster University for their hospitality during his
visit.  This work is supported by means of Natural Sciences and
Engineering Research Council of Canada Operating Grant OGP0008672 and
NATO Collaborative Research Grant 5-2-05/RG 910878.

\newpage

\begin{table}
\caption{Values for the shape parameter of the Reid hard-core
potential, calculated by different authors. $P_e$ and $P_b$ are shape
parameters determined from the eigenphases and `bar' phase shifts
respectively.  Values in parentheses are not given by the authors
but have been calculated from the value given by the author by using
the formula for the difference between the eigenphase and `bar' phase
shift shape parameter (see Eq. (\ref{2})).}
\begin{tabular}{rrc}
\multicolumn{1}{c}{$P_e$} &
\multicolumn{1}{c}{$P_b$} &
\multicolumn{1}{c}{Reference} \\   \hline
$(-0.015)$   &  $-0.011$   & \cite{reid68}        \\
$(-0.0109)$  &  $-0.0074$  & \cite{kermode81}     \\
$-0.0086$    &  $-0.0057$  & \cite{kermode91a}     \\
$0.006$      &  $(0.010)$  & \cite{wong92}        \\
$-0.00730$   &  $-0.00378$ & \cite{vandijk92}

\end{tabular}
\label{table1}
\end{table}

\begin{table}
\caption{Effective-range parameters for certain potentials.  The Reid
hard-core potentials used have three different cut-offs, {\it i.e.}, for Rhc1
$R = 14.285714 \mbox{ fm} + c$ (this is Reid's
potential\cite{reid68}),
for Rhc2 $R = 21.428571 \mbox{ fm} +
c$, and for Rhc3 $R =  28.571429 \mbox{ fm} + c$ where $c$ is the
hard-core radius.  Rsc is the Reid soft-core potential\cite{reid68}
and Moscow is the potential given in reference \cite{kukulin89}. }
\begin{tabular}{lllllllll}
\multicolumn{1}{c}{Potential} &
\multicolumn{1}{c}{$a_\alpha$ (fm)} &
\multicolumn{1}{c}{$r_\alpha$ (fm)} &
\multicolumn{1}{c}{$q$ (fm$^2$)} &
\multicolumn{1}{c}{$q_1$ (fm$^4$)} &
\multicolumn{1}{c}{$P_\alpha$} &
\multicolumn{1}{c}{$\bar P_1$} &
\multicolumn{1}{c}{$Q_\alpha$} &
\multicolumn{1}{c}{$\bar Q_1$}   \\ \hline
Rhc1 & 5.39713 & 1.72493 & 0.312461 & -2.13545 &
-0.007303 & -0.003779 & 0.0502 & 0.0719   \\
Rhc2 & 5.39701 & 1.72507 & 0.310400 & -2.07658 &
-0.007592 & -0.004115 & 0.0474 & 0.0684   \\
Rhc3 & 5.39701 & 1.72507 & 0.310356 & -2.07289 &
-0.007571 & -0.004094 & 0.0474 & 0.0684   \\
Rsc    & 5.38989 & 1.72225 & 0.316925 & -2.16703 &
-0.019085 & -0.015437 & 0.0483 & 0.0708  \\
Moscow & 5.39615 & 1.72266 & 0.326004 & -2.20509 &
-0.005972 & -0.002119 & 0.0516 & 0.0752  \\
\end{tabular}
\label{table2}
\end{table}

\end{document}